\documentstyle[aps,epsf,multicol]{revtex}
\begin{document}

\draft

\title{Explicit Green's Function of a Boundary Value Problem for a Sphere \\
and Trapped Flux Analysis in Gravity Probe B Experiment}

\author{I.M. Nemenman\thanks{Permanent Address: Department of Physics, Princeton University, Princeton, NJ 08544, USA} \and A.S. Silbergleit}
\address{Gravity Probe B, W.W.Hansen Experimental Physics Laboratory, 
\\Stanford University, Stanford, CA 94305-4085, USA}
\date{\today}
\maketitle

\begin{abstract}
Magnetic flux trapped on the surface of superconducting rotors of the Gravity Probe B (GP-B) experiment produces some signal in the SQUID readout. For the needs of GP-B error analysis and simulation of data reduction, this signal is calculated and analyzed in the paper. We first solve a magnetostatic problem for a point source (fluxon) on the surface of a sphere, finding the closed form elementary expression for the corresponding Green's function. Second, we calculate the flux through the pick-up loop as a function of the fluxon position. Next, the time dependence of a fluxon position, caused by rotor motion according to a symmetric top model, and thus the time signature of the flux are determined, and the spectrum of the trapped flux signal is analyzed. Finally, a multi-purpose program of trapped flux signal generation based on the above results is described,  various examples of the signal obtained by means of this program are given, and their features are discussed.
\end{abstract}

 
\section{Introduction}

The Gravity Probe B (GP-B) satellite is scheduled to fly in the year 2000. It contains a set of gyroscopes intended to test the predictions of general relativity that a gyroscope in a low (altitude$\approx650\,km$) circular polar orbit will precess, relative to a distant star, about 6.6 {\it arcsec/year} in the orbital plane (DeSitter, or geodetic, precession) and about 42 {\it milliarcsec/year} perpendicular to the orbital plane (Lense--Thirring, or frame--dragging, precession). To provide the desired measurement accuracy (1 part in $10^5$ for the geodetic effect), a magnetic London moment readout using SQUID has been chosen, so that the experiment will be carried out at low temperature ($\sim\, 2.5^\circ\,K$), 
and the gyro rotors will be superconducting (see~\cite{turn},~\cite{buch},~\cite{muhl} for the design  and  status of the experiment; the history of GP-B development is found in~\cite{fr}, and a survey of space relativity tests is in~\cite{mac}). The direction of the magnetic London moment developed in a rotating superconductor coincides with the direction of the rotation (spin) axis (F.London~\cite{lond}; for basic superconductor physics see~\cite{tink}; the description of gyromagnetic effects can be found in~\cite{ll8}, Ch. 4). The corresponding magnetic flux through the pick-up loop of the SQIUD is proportional to the sine of the angle between the London moment vector and the pick-up loop plane, so the change of this angle, and thus the drift of the gyroscope axis, can be detected from the SQUID signal at the roll frequency of the spacecraft which will be deliberately rotated.

However, along with the London moment dipole, there will also be quantum--size sources of magnetic field (fluxons) pinned to the surface of the superconducting rotor (see~\cite{tink}, Ch. 5; ~\cite{rir}, Ch. 12) which  produce additional magnetic flux through the pick-up loop called trapped flux; its time signature will be present in the SQUID output. The low frequency part of this signal, though comparatively small under the GP-B conditions, might corrupt the accuracy of the London moment readout. On the other hand, its high frequency part can provide additional information significant for the experimental results. To make sure the trapped flux does not affect the measurement precision, as well as to extract useful information from it, one has to analyze the trapped flux signal and develop the code generating it, for the use in simulations of the GP-B  error analysis and data reduction. This is the aim of the present paper. Note that the first work on the analysis of the trapped flux from a GP-B rotor was done by L.Wai in his thesis~\cite{wai}.

In sec. II we give a closed form solution to a magnetostatic problem of a fluxon on the surface of the gyroscope. In sec. III the solution is used to find the trapped flux signal in the pick-up loop as a function of the fluxon's position. The closed form expression for the trapped flux appears to be not very useful for further applications, so various exact and approximate formulas are also obtained. In sec. IV we investigate the motion of a fluxon with respect to the pick-up loop, thus finding the time signature of the trapped flux signal; we then go on to analyze its frequency spectrum. The last section contains a brief description of the program used to simulate trapped flux for the GP-B data processing routines. Pictures of the high frequency signal, its low frequency envelope, and various Fourier spectra are presented and discussed.

\section{Green's function of the magnetostatic problem} 

The GP-B experiment will be conducted at low temperatures, so the fluxons can be treated as static (welded to the rotor's surface) and non-interacting ones. In such a case the total fluxon field is a superposition of the fields of individual fluxons. In addition, the rate of change of this field due to the rotor's motion is negligible, hence the magnetostatic approach should be used. Thus we consider a single fluxon whose characteristic size is on the order of $10^{-5}\,cm$ (~\cite{rir}, p. 184); due to a macroscopic size of the gyroscope (1.91 {\it cm} radius), the fluxon can be treated as a point source of magnetic field with the coordinate angles $\vartheta_f,\,\varphi_f$ on the surface $r=r_g$ of the rotor. The spherical coordinates $r,\vartheta,\,\varphi$ here correspond to a Cartesian frame $\{x,y,z\}$ fastened to the pick--up loop so that the origin coincides with the loop center and the $z$ axis is perpendicular to the loop plane; the real relative motion of the fluxon and the loop, i. e., the dependence of the fluxon position angles $\vartheta_f,\,\varphi_f$ on time, will be incorporated and examined in sec. IV.

In these settings, the boundary value problem for the magnetic potential $\Psi({\bf r})$ of the fluxon outside the rotor is formulated as

\begin{equation}
\Delta \Psi \left({\bf r} \right) = 0,  \qquad r > r_g,\,\, 0 \leq \theta_f\leq\pi,\,\, 0 \leq \varphi_f <2\pi
\label{eq:laplace}	 
\end{equation}
	
\begin{equation}
 -\frac{\partial \Psi}{\partial r}\biggl|_{r=r_g} = \frac{\Phi_0}{r^2_g {\rm sin} \vartheta_f} \delta \left(\vartheta - \vartheta_f \right) \delta \left( \varphi - \varphi_f \right), 
\label{eq:bvp-Neumann}
\end{equation}
where $\Phi_0=h/2e$ is the magnetic flux quantum, and the magnetic field is
\begin{equation}
{\bf B}= -{\bf \nabla} \Psi
\label{eq:magnfield}
\end{equation}
Evidently, up to a factor $\Phi_0$, $\Psi$ is the Green's function of the external Neumann boundary value problem for a sphere.

A standard separation of variables leads to the following series representation of the solution to (\ref{eq:laplace}), (\ref{eq:bvp-Neumann}):
\begin{equation}
\Psi \left( {\bf r} \right) \equiv \Psi\left(r,\vartheta,\varphi \right)= \frac{\Phi_0}{2\pi r_g} \sum_{l=0}\limits^{\infty} \sum_{m=0}\limits^{l} \left(M_{lm} \cos m\varphi + N_{lm} \sin m \varphi \right) \left( \frac{r_g}{r} \right)^{l+1} P^m_l \left(\cos\vartheta \right),
\label{eq:psi-separation}
\end{equation}
with the coefficients given by
\begin{equation}
M_{lm} = \frac{2l+1}{\left(1+\delta_{m0} \right) \left( l+1 \right)} \frac{\left( l-m \right)!}{\left( l+m \right)! }
 P^m_l \left( {\cos} \vartheta_f \right)\cos m\varphi_f, \qquad
N_{lm} = \frac{2l+1}{\left( l+1 \right)} \frac{\left( l-m \right)!}{\left( l+m \right)! }
P^m_n \left( \cos\vartheta_f \right) \sin m \varphi_f
\label{eq:psi-coeffs}
\end{equation} 

As it turns out, this series may be summed to give the closed form solution for $\Psi$. To determine it, we first introduce (\ref{eq:psi-coeffs}) into (\ref{eq:psi-separation}) to obtain
$$
\Psi \left( {\bf r} \right)= \frac{\Phi_0}{4\pi r_g} \sum_{l=0}\limits^{\infty}\frac{2l+1}{l+1}\left( \frac{r_g}{r} \right)^{l+1}
\left[
P_l \left(\cos\vartheta \right)P_l \left(\cos\vartheta_f \right)+
2\sum_{m=0}\limits^{l}  P^m_l \left(\cos\vartheta \right)P^m_l \left(\cos\vartheta_f \right)\cos m\left(\varphi-
\varphi_f\right)
\right]
$$
Then, by applying the addition theorem for Legendre functions (see~\cite{bat2}, 10.11, (47)), we convert the latter into 
\begin{equation}
\Psi\left({\bf r}\right) = \frac{\Phi_0}{4 \pi r_g} \sum_{l=0}^{\infty} \frac{2l+1}{l+1}\left(\frac{r_g}{r} \right)^{l+1} P_l \left(\cos\gamma \right)= 
\frac{\Phi_0}{4 \pi r_g}
\left[
 2\sum_{l=0}^{\infty}\left(\frac{r_g}{r} \right)^{l+1} P_l \left(\cos\gamma \right)-
\sum_{l=0}^{\infty}\frac{1}{l+1}\left(\frac{r_g}{r} \right)^{l+1} P_l \left(\cos\gamma \right)
\right],
\label{eq:psi-singleseries}
\end{equation}
where $\gamma$ is the angle between the directions to the fluxon and to the observer:
\begin{equation}
{\cos} \gamma \equiv {\rm cos} \vartheta {\rm cos} \vartheta_f + {\rm sin} \vartheta {\rm sin} \vartheta_f {\rm cos} \left( \varphi - \varphi_f \right) 
\label{eq:gamma}
\end{equation}

The first of the series in the above expression for $\Psi$ is obviously the generating function for Legendre polynomials (see~\cite{bat2}, 10.10, (39)), the second one is just an integral of it, namely,
$$
\sum_{l=0}^{\infty}\frac{1}{l+1}\eta^{l+1} P_l \left(\zeta\right)=
\int\limits_0^\eta d\tau \sum_{l=0}^{\infty}\tau^{l} P_l \left(\zeta \right)=
\int\limits_0^\eta \frac{d\tau}{\sqrt{1-2\zeta\tau+\tau^2}}=
\ln\frac{\eta-\zeta+\sqrt{1-2\zeta\eta+\eta^2}}{1-\zeta} 
$$
Using these results in (\ref{eq:psi-singleseries}), we can now write the magnetic potential in its final form as a finite combination of elementary functions:
\begin{equation}
\Psi\left({\bf r}\right)\equiv\Phi_0\,G({\bf r},{\bf r_f}) = 
\frac{\Phi_0}{2\pi} \left[ \frac{1}{\left| {\bf r} - {\bf r_f}\right|} - \frac{1}{2r_g} \ln\frac{ r_g^2 - {\bf r} \cdot {\bf r_f} +r_g \left| {\bf r} - {\bf r_f} \right|}{ rr_g -{\bf r} \cdot {\bf r_f}}\right],
\label{eq:psi-closed}
\end{equation} 
where $G({\bf r},{\bf r_f})$ is the mentioned Green function and ${\bf r_f}=\{r_g,\vartheta_f,\,\varphi_f\}$ is the position vector of the source. The first term here, as one would expect, is a half of the potential of a point charge, and the addition to it describes the contribution of the curved boundary.

Since, surprisingly enough, we were not able to find this explicit formula in literature, it seems reasonable to give here a closed form expression for the Green function of the corresponding Dirichlet problem ($G_D$), in which the boundary condition (\ref{eq:bvp-Neumann}) is replaced by
\begin{equation}
\Psi|_{r=r_g} = \frac{\Phi_0}{r_g \sin \vartheta_f} \delta \left(\vartheta - \vartheta_f \right) \delta \left( \varphi - \varphi_f \right)
\label{eq:bvp-Dirichlet}
\end{equation}
The result then is
\begin{equation}
\Psi\left({\bf r}\right)\equiv\Phi_0\,G_D({\bf r},{\bf r_f})=\frac{\Phi_0}{4\pi} \frac{r^2 -r_g^2}{\left| {\bf r} -{\bf r_f} \right|^{3}}
\label{eq:bvp-Dclosed}
\end{equation}
Note that Green's functions for the corresponding internal problems can be obtained from~(\ref{eq:psi-closed}) and~(\ref{eq:bvp-Dclosed}) by means of inversion.

\section{Trapped Flux as a function of a fluxon position}
Magnetic flux measured by the pick-up loop of a GP-B SQUID is the flux through the circle of the radius $R$ in the plane $z=0$, or, equivalently, the flux through the (upper) hemisphere. The dependence of the trapped flux on the fluxon position turns out to be rather complicated, especially for the GP-B design, when the gap between the rotor and the loop is very small as compared to the pick-up loop radius $R$. For that reason we give here a number of different representations of the trapped flux as a function of the fluxon position; each of them has its own merits and drawbacks and is thus used for different purposes pertinent to our investigation.

\subsection{Trapped flux in terms of series of Legendre polynomials}

The simplest way to calculate the trapped flux is to integrate over the hemisphere the series expression for the radial component of the magnetic field obtained from~(\ref{eq:magnfield})--(\ref{eq:psi-coeffs}):
$$
\Phi_f=\int\limits_{hemisphere\, (r=R)}\,B_r\biggl|_{r=R}\,dA=
\int\limits_{hemisphere\, (r=R)}\,-{\partial\Psi\over\partial r}\biggl|_{r=R}\,dA=
\Phi_0\sum_{l=0}^\infty (l+1)\Bigl(
{r_g\over R}
\Bigr)^{l}M_{l0}\int_0^1\,P_l(s)\,ds;
$$	
all spherical harmonics with $m\not=0$ here have averaged out over the azimuthal angle $\varphi$. The last integral is calculated with the help of the known relations of the theory of Legendre polynomials (see~\cite{bat2}, 10.10, (14), (2), (4)) :
$$
P_l(s)=\frac{P_{l+1}^{\prime}(s)-P_{l-1}^{\prime}(s)}{l+1};\quad P_l(1)=1;\quad P_{2k+1}(0)=0;
\quad P_{2k}(0)={(-1)^{k}\over \sqrt{\pi}}\,{\Gamma(k+1/2)\over k!};\quad l,k=0,1,\dots;
$$
$\Gamma(\zeta)$ is the Euler gamma-function. Then, after inserting the values $M_{l0}$ from~(\ref{eq:psi-coeffs}), we arrive at the following expressions:
$$
\Phi_f(\cos\vartheta_f)=\frac{\Phi_0}{2}F_\delta(\cos\vartheta_f);
$$
\begin{equation}
F_\delta(s)=\sum_{k=0}^\infty\left(1-\delta\right)^{2k+1}P_{2k+1}(s)\left[P_{2k}(0)-P_{2k+2}(0)\right] =
{2\over \sqrt{\pi}}\sum_{k=0}^\infty (-1)^{k}\,{k+3/4\over (k+1)!}\,\Gamma(k+1/2)\,
\left(1-\delta\right)^{2k+1}P_{2k+1}(s)
\label{eq:univ_series}
\end{equation} 
Here $\delta$ denotes the dimensionless gap between the pick-up loop and the rotor, $0\leq\delta=(R-r_g)/R<1$.

From the point of view of signal processing, $F_\delta(s)$ is a transfer function which converts the "input" fluxon position signal $S_{in}(t)=\cos\vartheta_f(t)$ (the position is changing with the time as the rotor moves relative to the pick-up loop, see the next section), into an "output" trapped flux signal $S_{out}(t)=0.5\Phi_0F_\delta(S_{in}(t))$ which is present in the GP-B readout. For the reason that the total contribution to the flux of any number of fluxons scattered in any way over the rotor's surface is given by the sum of the values of the same function $F_\delta$ taken at proper different values of its argument, it was called "universal curve" in~\cite{wai}. Clearly, $F_\delta(s)$ is an odd function of $s$; in particular, $F_\delta(0)=0$ means that a fluxon sitting exactly in the pick-up loop plane does not, of course, register any flux.

By setting $\delta=0$ in~(\ref{eq:univ_series}) (the loop on the surface of the rotor), we immediately find  
\begin{equation}
F_0(s)={2\over \sqrt{\pi}}\sum_{k=0}^\infty (-1)^{k}\,{k+3/4\over (k+1)!}\Gamma(k+1/2)P_{2k+1}(s)=
\cases{1&if $\,0<s\leq 1$;\cr 0&if $\,s=0$;\cr -1&if $\,-1\leq s<0$.\cr}
\label{eq:univ_d0}
\end{equation}
(the last equality here is proved by expanding its right-hand side in orthogonal series of Legendre polynomials).

This result obtained by L.Wai~\cite{wai} has a clear physical meaning: when the pick-up loop lies on the rotor's surface, same as the point source of field always does, the flux through the loop remains unchanged ($\pm{\Phi_0}/{2}$, half of the total) while the fluxon stays in either of the hemispheres separated by the plane of the loop,
and changes it sign by a jump when the fluxon crosses this plane. However, equation~(\ref{eq:univ_d0}) also demonstrates the difficulties in using expression~(\ref{eq:univ_series}) for GP-B,  where $\delta=0.025$ is very small: for any  $\delta>0$ the series~(\ref{eq:univ_series}) has an absolutely converging majorant, so its sum $F_\delta(s)$ is an analytical function of $s$, but it has a jump discontinuity at $s=0$ when $\delta=0$. Therefore the series~(\ref{eq:univ_series}) converges worse and worse with the separation  $\delta$ becoming smaller and smaller, which makes~(\ref{eq:univ_series}) practically unacceptable for accurate numerical calculations at the required value of separation. It also turns finding {\it a uniform in $s$} asymptotic expansion of $F_\delta(s)$ for $\delta\rightarrow0$ into a rather difficult mathematical problem. The effect is that for small positive values of $\delta$ the transfer function has a shape of a very steep "kink" (recall that $F_\delta(s)$ is odd): it is almost constant outside a small vicinity $(-\Delta_\delta,\,\Delta_\delta)$ of the origin, with $\Delta_\delta=\rm{O}(\delta)$ as shown below, and is equal to zero at $s=0$ with a huge gradient $\sim\rm{O}(1/\delta)$ there (see fig. 1). That is why we are deriving three more representations for $F_\delta(s)$ in the following subsections.

\subsection{Integral representation of the trapped flux}

An integral expression for $F_\delta(s)$ is obtained by replacing the Legendre polynomials in~(\ref{eq:univ_series}) by their integral representation (see~\cite{bat2}, 10.10, (43))
$$
P_{2k+1}(\cos\vartheta_f)=\frac{1}{\pi}\int_
{-\vartheta_f}^{\vartheta_f}\frac{\exp\left[i\left(2k+1+1/2\right)\right]\,d\psi}{\sqrt{2\left(\cos\psi-\cos\vartheta_f\right)}}
$$
Changing then the order of summation and integration, we arrive at a sum of two hypergeometric series 
which are readily summed up to result in:
\begin{equation}
F_\delta(\cos\vartheta_f)=\frac{\Phi_0 \sqrt{2}}{\pi} \int_{0}^{\vartheta_f}
\frac{d\psi \exp \left( i\psi/2 \right)}{\sqrt{\cos \psi -\cos \vartheta_f}}
\left[ \frac{\lambda}{\sqrt{1+\lambda^2}}- \frac{\sqrt{1+\lambda^2}}{2\lambda} + \frac{1}{2\lambda} \right]
,\,\,\,\, \lambda\equiv\left(1-\delta\right)\exp \left(i\psi\right)
\label{eq:univ_integr}
\end{equation}

Representation~(\ref{eq:univ_integr}) is very convenient for precise numerical calculation (and, in fact, is used for this purpose in our code, see sec. V), because the integrand in~(\ref{eq:univ_integr}) is an algebraic one, and the weak singularity at the upper limit can be taken care of rather easily. The plot of the transfer function computed from~(\ref{eq:univ_integr}) is given in fig.~1, along with the graphs of its various approximations described in the next subsection. The relative error of the numerical computation has been kept within $10^{-5}$.

\subsection{Elementary approximations of the trapped flux}

From the described behavior of $F_\delta(s)$ for small $\delta$ it is clear that to effectively approximate it one  needs the value of its gradient at $s=0$ and the "saturation" value $F_\delta(1)$, in the first place. Fortunately, it is possible to compute these quantities exactly, and they are
\begin{equation}
f_\delta\equiv F_\delta(1)={1\over 1-\delta}\,
\left[
1-{2\delta-\delta ^2\over\sqrt{1+(1-\delta)^2}}
\right]=1-(\sqrt{2}-1)\delta+O(\delta ^2);
\label{eq:fd}
\end{equation}
\begin{equation}
\kappa_\delta \equiv{\partial F_\delta(s)\over\partial s}\biggl|_{s=0}=
{2\over\pi}\,
\left[
{1+(1-\delta)^2\over 1-(1-\delta)^2}\,{\bf E}(1-\delta )-{\bf K}(1-\delta )
\right]=
{2\over\pi}\,
\left[
{1\over\delta} +2+O(\delta\log\delta^{-1})
\right], \quad \delta\rightarrow0;
\label{eq:kd}
\end{equation}
here ${\bf K}(k),\,{\bf E}(k)$ are complete elliptic integrals of the first and second kind, respectively (see~\cite{dw}, Ch. IX for their definitions and asymptotic behavior at $k\rightarrow1-0$). The formulas are derived from~(\ref{eq:univ_series}) by the direct summation of the corresponding series of Legendre polynomials carried out in the Appendix.

The simplest approximation of the transfer function for $\delta\rightarrow+0$ is evidently a piecewise-linear one,
\begin{equation}
F_\delta(s)\approx\cases{1&if $\,\Delta_\delta<s\leq 1$;\cr \kappa_\delta s&if $\,|s|\leq\Delta_\delta$;\cr -1&if $\,-1\leq s<-\Delta_\delta$,\cr}
\label{eq:univ_lin}
\end{equation}
with $\Delta_\delta$ defined in a natural way as
\begin{equation}
\kappa_\delta\,\Delta_\delta=f_\delta, \qquad\Delta_\delta={f_\delta\over\kappa_\delta}={\pi\over2}\,\delta+\rm{O}(\delta^2)
\label{eq:Dd}
\end{equation}
It turns out that this approximation gives the right qualitative picture of the signal and even is not too bad quantitatively, providing, for all values $|s|\leq1$, the error within $1/3$ for both $\delta=0.3$ and $\delta=0.025$. This accuracy, however, is not enough for the GP-B simulations, moreover, the largest error, associated with the jump of the derivative of function~(\ref{eq:univ_lin}) at $s=\pm\Delta_\delta$, occurs in a very sensitive transition region where the fast growth of $F_\delta(s)$ is replaced by its almost constant behavior.

A much more attractive approximation is given by the function
\begin{equation}
F_\delta(s)\approx{2\over\pi}f_\delta\,\arctan\left({\pi\over2}{\kappa_\delta s\over f_\delta}\right),
\qquad \delta\rightarrow+0
\label{eq:univ_atan}
\end{equation}
The parameters here are arranged in such a way that the slope at $s=0$ is exactly $\kappa_\delta$ and, in the spirit of asymptotic methods, the true saturation value is achieved when $\kappa_\delta s=\infty$ (note that another "simple and natural" approximating function, the hyperbolic tangent, is not acceptable, because the rate of approaching of $f_\delta$ by $F_\delta(s)$ is a power rather than exponential one).  The performance of the approximation~(\ref{eq:univ_atan}) exceeds all expectations, giving, over the whole range of $s$, the maximum error of $20\%$ for $\delta=0.3$, and only $1.8\%$ for $\delta=0.025$. 
\vskip2mm
\centerline{\epsfxsize=0.6\hsize\epsffile{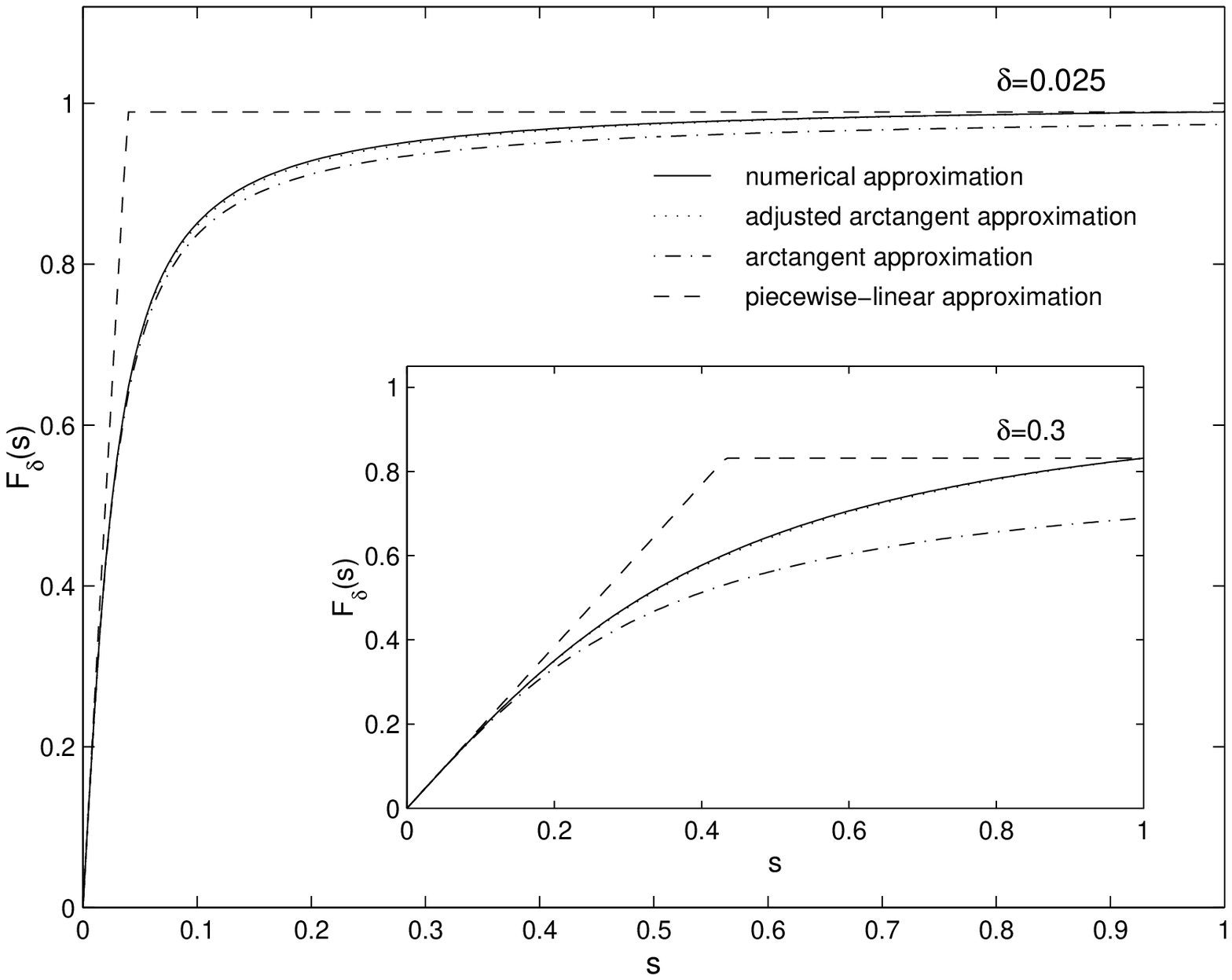}}
\vskip1mm
\centerline{Fig. 1. Universal Curve $F_{\delta}(s)$.}
\vskip2mm
The accuracy is mostly lost outside the transition zone $(-\Delta_\delta,\,\Delta_\delta)$ due to the fact that $f_\delta$ is achieved only at infinity. This can be dealt with by redefining the parameters to have both the exact slope at $s=0$ and the right value at $s=1$, which produces 
\begin{equation}
F_\delta(s)\approx A_\delta\,\arctan{\kappa_\delta s\over A_\delta},\qquad
A_\delta\,\arctan{\kappa_\delta \over A_\delta}=f_\delta,
\qquad \delta\rightarrow+0
\label{eq:univ_atan1}
\end{equation}
This `adjusted' arctan gives the maximum error within $0.3\%$ for $\delta=0.025$; and even for as large a separation as $\delta=0.3$ the error is still about $0.6\%$. Same as~(\ref{eq:univ_lin}) and~(\ref{eq:univ_atan}), the dependence~(\ref{eq:univ_atan1}) is shown in fig. 1. $A_\delta$  versus $\delta$ is plotted in fig. 2; note a relative flatness of the of the function. 
\vskip2mm
\centerline{\epsfxsize=0.6\hsize\epsffile{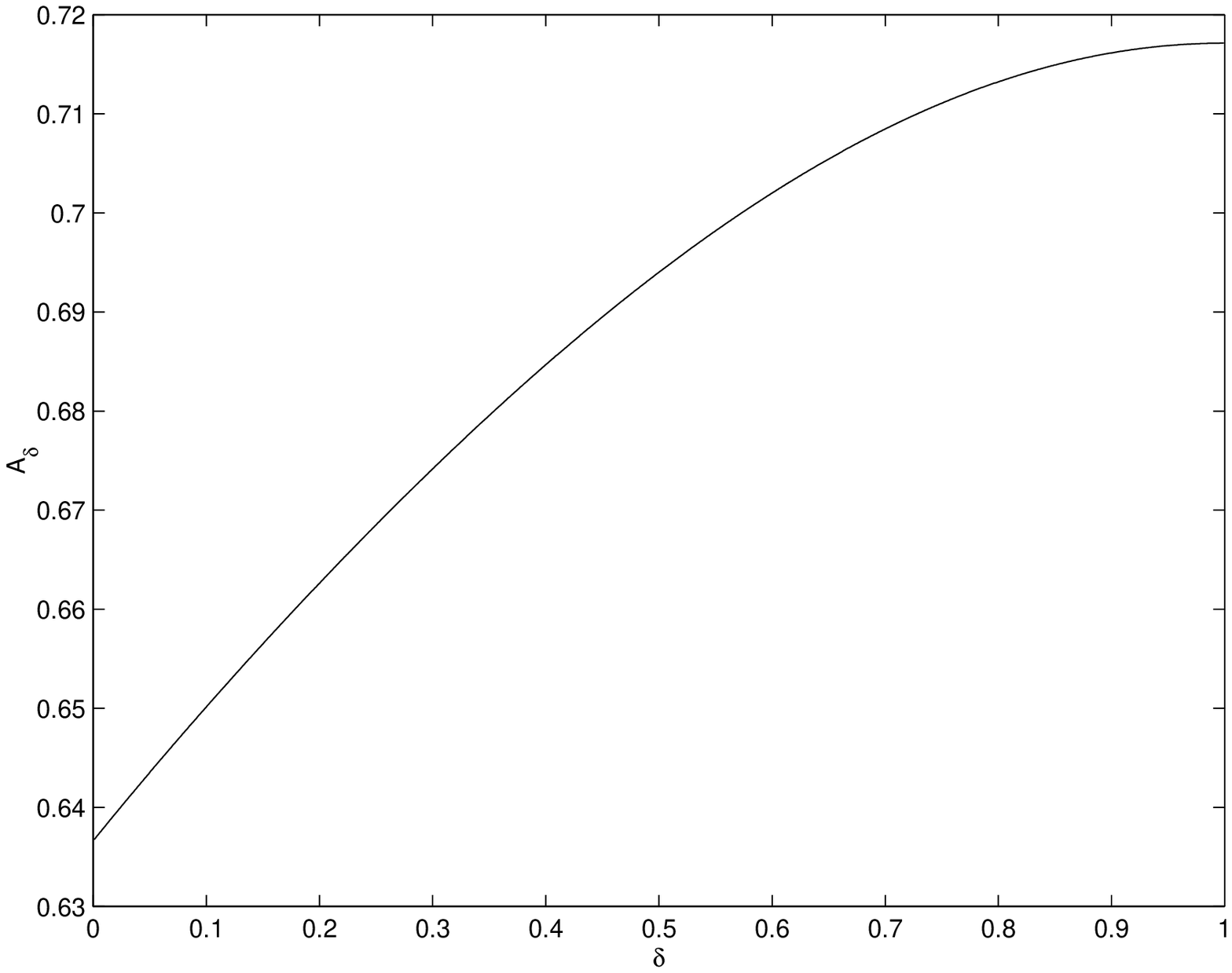}}
\vskip1mm
\centerline{Fig. 2. Dependence of $A_{\delta}$ on $\delta$.}
\vskip2mm

\subsection{Closed form expression of the trapped flux}

The explicit formula for the trapped flux can also be obtained, though not that easily, from equation~(\ref{eq:univ_series}), however, a direct way to get it is to integrate the closed form expression for the magnetic field through the pick-up loop plane $z=0$. For this plane $\vartheta=\pi/2$, $r=\rho$ (the polar radius); in addition, we can redefine $\varphi$ by setting $\varphi_f=0$. Then equations~(\ref{eq:psi-closed}) and~(\ref{eq:magnfield}) provide the needed component of the magnetic field in the form:
\begin{equation}
B_z\biggl|_{z=0} = -\frac{\Phi_0 r_g \cos \vartheta_f}{2 \pi}
\left[
\frac{1}{X^3(\rho,\varphi)}+
\frac{\rho-r_g\sin \vartheta_f \cos \varphi}{2r_g^2\rho X(\rho,\varphi)\, Y_+(\varphi)\, Y_-(\varphi)}+
\frac{\sin \vartheta_f \cos \varphi}{2r_g^2 \rho \, Y_+(\varphi)\, Y_-(\varphi)}-
\frac{1}{2r_g^2\rho \, Y_-(\varphi)}
\right]
\label{eq:magnfield_z0}
\end{equation}
where
\begin{equation}
X(\rho, \varphi) =  \sqrt{ r_g^2 - 2r_g\rho \sin \vartheta_f \cos \varphi +\rho^2 },\,\,\,\,\,
Y_{\pm}(\varphi) = 1 \pm \sin \vartheta_f \cos \varphi
\label{eq:XY}
\end{equation}
Now we need to integrate~(\ref{eq:magnfield_z0}) over the area of the pick-up loop. First we calculate the simple, though rather cumbersome, algebraic integral of the field~(\ref{eq:magnfield_z0}) times $\rho d\rho$ over the polar radius from $0$ to $R$ (if instead one first integrates over $\varphi$, elliptic integrals of a complicated argument appear in the result that make the closed form radial integration very difficult). As we are then to integrate over the period of $\cos\varphi$, the terms {\it odd} in $\cos \varphi$ can be omitted, and we obtain:
\begin{equation}
\Phi_f(\cos \vartheta_f) = \frac{\Phi_0}{2}F_{\delta}(\cos \vartheta_f)=
-\frac{\Phi_0 r_g \, \cos \vartheta_f}{2\pi}\int_0^{2\pi}d\varphi
\left[
\frac{R^2-r_g^2}{2r_g^2\, X(R,\varphi) \, Y_+(\varphi) \, Y_-(\varphi)}-
\frac{R}{2r_g^2\, Y_-(\varphi)}
\right]
\label{eq:flux_1int}
\end{equation} 
In view of~(\ref{eq:XY}), this integration is also rather straightforward and leads to the desired result:
\begin{equation}
\Phi_f(\cos \vartheta_f)=\frac{\Phi_0}{2}\frac{\cos \vartheta_f}{1-\delta}
\left\{
\frac{1}{\left| \cos \vartheta_f \right|}-
\frac{2\delta -\delta^2}{\pi \, \sqrt{2(1-\delta)(1+\sin \vartheta_f)+\delta^2}}
\left[\frac{{\bf \Pi}(\nu_+,k)}{1+\sin \vartheta_f} + \frac{{\bf \Pi}(\nu_-,k)}{1-\sin \vartheta_f}\right]
\right\},
\label{eq:flux_closed}
\end{equation} 
where
\begin{equation}
\nu_{\pm}(\vartheta_f) = \mp \frac{2\sin\vartheta_f}{1 \pm \sin\vartheta_f},\qquad
k(\vartheta_f,\delta)= \sqrt{\frac{4\left(1-\delta \right) \sin \vartheta_f}
{2\left(1-\delta\right)\left(1+\sin\vartheta_f\right)+\delta^2}}
\label{eq:Ppm}
\end{equation}
and ${\bf \Pi}(\nu,k)$ is the complete elliptic integral of the third kind (see~\cite{dw}, Ch.IX). As a consistency check, one may calculate the saturation value and the derivative at zero of the transfer function~(\ref{eq:flux_closed}) to see that they are indeed equal to the previously obtained values~(\ref{eq:fd}) and~(\ref{eq:kd}). 

The first term in~(\ref{eq:flux_closed}) evidently has a jump discontinuity at $s=\cos\vartheta_f=0$. Therefore, for all finite $\delta$, the second term must contain the discontinuity of the opposite sign, to make the sum of two analytical in $s$. Hence for small positive $\delta$ in the transition zone we are dealing with a small difference of two large quantities, which is always a problem. Also, the first term in~(\ref{eq:flux_closed}) coincides exactly with the expression (12) for $\delta=0$, hence the second one should disappear in this limit, which it necessarily does in a very nonuniform way. Evidently, such an expression cannot be effectively used for both numerical and analytical purposes when $\delta$ is small enough, which is our case.

\section{Fluxon kinematics and spectral decomposition of the trapped flux signal}

Now we need to determine the time signature $\vartheta_f(t)$ of a fluxon polar angle in the pick-up loop frame, to complete the investigation of the trapped flux signal.

In doing that we use four Cartesian coordinate systems. The first one, $\{x,\,y,\,z\}$, has been introduced in sec. I; it is fastened to the pick--up loop,
and ${\bf z}$  is the unit vector normal to the loop plane. The second coordinate system, $\{x_r,\,y_r,\,z_r\}$, is associated with the roll axis of the spacecraft, $\hat\omega_r={\bf z}_r$ (fig. 3).
The roll axis is almost in the pick-up loop plane, that is, the roll axis---pick-up loop plane misalignment $\alpha\leq10^{-5}$ is very small. The third set of coordinates, $\{x_L,\,y_L,\,z_L\}$, is related to the angular momentum vector ${\bf L}$ in a way that 
${\bf z}_L={\bf L}/|{\bf L}|$. Both $r$-- and $L$--coordinates are fixed in the inertial space, since the roll axis is pointed to the Guide Star, and we can so far neglect the pointing errors, as well as the relativistic drift of ${\bf L}$. We choose axes ${\bf y}_r$ and ${\bf y}_L$ in the plane containing both ${\bf z}_r$ and ${\bf z}_L$, then the perpendicular to this plane axes ${\bf x}_r$ and ${\bf x}_L$ coincide (fig. 4), and the following relations are true:
$$
{\bf z}_L\cdot{\bf z}_r={\bf y}_L\cdot{\bf y}_r=\cos\beta_0,\qquad
{\bf z}_L\cdot{\bf y}_r=-{\bf y}_L\cdot{\bf z}_r=\sin\beta_0
$$
\begin{equation}
{\bf x}_r\cdot{\bf z}_r={\bf x}_r\cdot{\bf z}_L={\bf x}_r\cdot{\bf y}_r={\bf x}_r\cdot{\bf y}_L=0
\label{eq:lrrelations}
\end{equation}
Here $\beta_0$ is the roll axis---angular momentum misalignment which is required to be $\leq 5\times10^{-5}rad$ in the GP-B experiment.

\vskip2mm
\centerline{\epsfxsize=0.6\hsize\epsffile{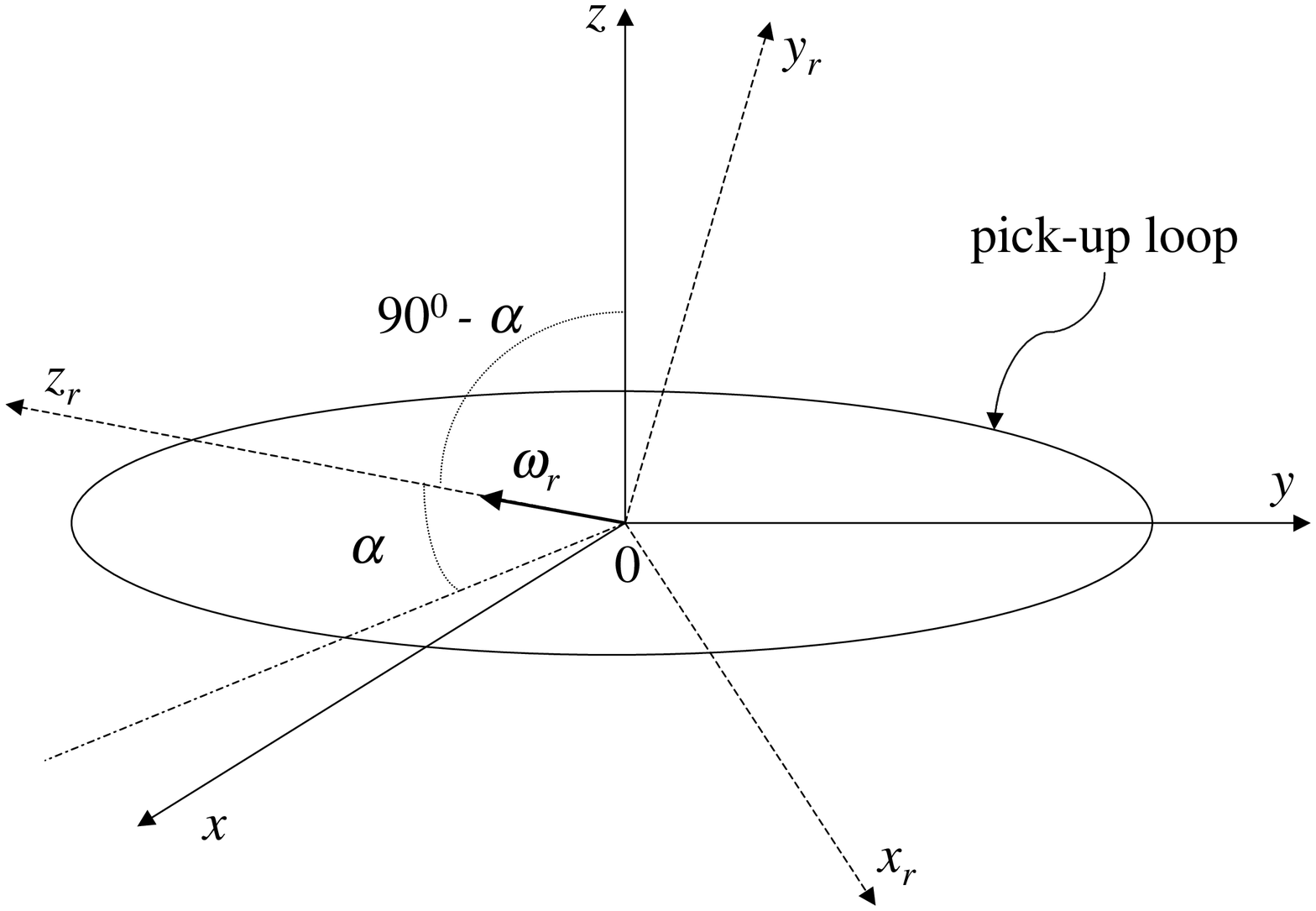}}
\centerline{Fig. 3. Mutual Orientation of Roll and Loop Coordinates.}

\centerline{\epsfxsize=0.6\hsize\epsffile{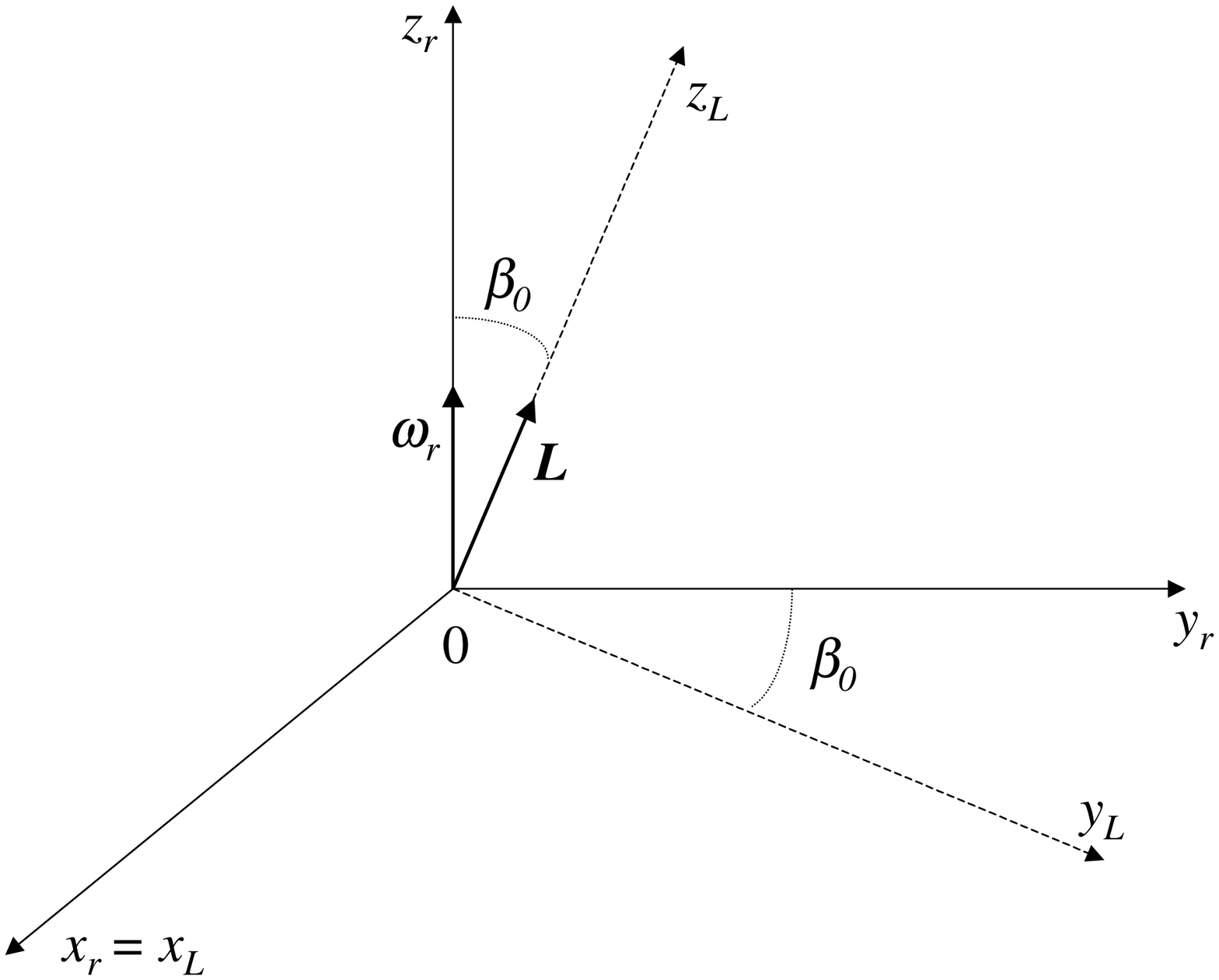}}
\centerline{Fig. 4. Mutual Orientation of Roll and Angular Momentum Coordinates.}
\vskip2mm

A  symmetric top with the moment of inertia $I+\Delta I$ relative to the body symmetry axis  and equal and slightly different value $I$ for the moments of inertia about the other two axes is a very good model for the GP-B rotors (note that ${|\Delta I|/I}\leq 10^{-5}$ for them). Therefore, we choose the fourth Cartesian coordinate system $\{{x}_B,\,{y}_B,\,{z}_B\}$ fixed in the rotor's body with ${\bf z}_B$ directed along the rotor's symmetry axis.

The dynamics of a symmetric rotor is well known and relatively simple (c. f.~\cite{landau,golds}). Its motion in the $L$-coordinates is a precession about ${\bf z}_L$ with the spin frequency
\begin{equation}
\omega_s={L \over I},
\label{eq:ws}
\end{equation}
and rotation about the rotor symmetry axis ${\bf z}_B$ with the frequency
\begin{equation}
\omega_{rot}={L \over I+\Delta I}\cos\gamma_B\simeq\omega_s
\Bigl(1-{\Delta I \over I}\Bigr)\cos\gamma_B;
\label{eq:wrot}
\end{equation}
$0\leq\gamma_B\leq\pi$ is the angle between ${\bf z}_L$ and ${\bf z}_B$.

For the signal of the trapped field we need, however, the time dependence of the position of a fluxon in the inertial coordinates, hence we need expressions of ${\bf x}_B(t),\,{\bf y}_B(t),\,{\bf z}_B(t)$ in terms of ${\bf x}_L,\,{\bf y}_L,\,{\bf z}_L$. The latter is found with the help of the Euler angles (see for instance~\cite{landau}) in the form
$$
\noindent {\bf z}_B(t)={\bf z}_L\,\cos\gamma_B+{\bf x}_L\,\sin\gamma_B\cos\theta_s+{\bf y}_L\,\sin\gamma_B\sin\theta_s
$$
$$
{\bf y}_B(t)=-{\bf z}_L\,\sin\gamma_B\cos\theta_p+
$$
$$
{\bf x}_L\,\bigl(\cos\gamma_B\cos\theta_s\cos\theta_p-\sin\theta_s\sin\theta_p\bigr)+{\bf y}_L\,\bigl(\cos\gamma_B\sin\theta_s\cos\theta_p+\cos\theta_s\sin\theta_p \bigr)
$$
$$
{\bf x}_B(t)=-{\bf z}_L\,\sin\gamma_B\sin\theta_p+
$$
\begin{equation}
{\bf x}_L\,\bigl(\cos\gamma_B\cos\theta_s\sin\theta_p+\sin\theta_s\cos\theta_p\bigr)+{\bf y}_L\,\bigl(\cos\gamma_B\sin\theta_s\sin\theta_p-\cos\theta_s\cos\theta_p\bigr)
\label{eq:BofL}
\end{equation}
Here the spin and polhode phases are
\begin{equation}
\theta_{s}(t)=\omega_{s}t+\theta_{s}^0,\qquad \theta_{p}(t)=\omega_{p}t+\theta_{p}^0,\qquad \theta_{s,p}^0=const,
\label{eq:phases}
\end{equation}
and $\omega_p$ is a polhode frequency,
\begin{equation}
\omega_p={L \over I}{|\Delta I| \over I}\cos\gamma_B=\omega_s{|\Delta I| \over I}\cos\gamma_B
\label{eq:wp}
\end{equation}
(In the body-fixed frame the instant angular velocity vector rotates around the rotor's symmetry axis with the polhode frequency). Using this, we obtain the following expression for the unit vector ${\bf e}_f$ in the direction of a fluxon (i. e., of an arbitrary fixed point of the rotor surface at some polar, $0\leq\xi\leq\pi$, and azimuthal, $0\leq\eta<2\pi$, angles in the body-fixed spherical coordinates):
$$
{\bf e}_f={\bf z}_B(t)\,\cos\xi+\bigl({\bf x}_B(t)\cos\eta+{\bf y}_B(t)\sin\eta\bigr)\sin\xi\equiv e_1(t){\bf x}_L+ e_2(t){\bf y}_L+e_3(t){\bf z}_L
$$
$$
e_1(t)=\sin\xi\Bigl[\cos\gamma_B\cos\theta_s(t)\sin\bigl( {\theta_p}(t)+\eta\bigr)+\sin\theta_s(t)\cos\bigl( {\theta_p}(t)+\eta\bigr)\Bigr]+\cos\xi\sin\gamma_B\cos\theta_s(t)
$$
$$
e_2(t)=\sin\xi\Bigl[\cos\gamma_B\sin\theta_s(t)\sin\bigl( {\theta_p}(t)+\eta\bigr)+\cos\theta_s(t)\cos\bigl( {\theta_p}(t)+\eta\bigr)\Bigr]+\cos\xi\sin\gamma_B\sin\theta_s(t)
$$
\begin{equation}
e_3(t)=-\sin\xi\sin\gamma_B\sin\bigl( {\theta_p}(t)+\eta\bigr)+\cos\xi\cos\gamma_B
\label{eq:ef}
\end{equation}

According to the results of sec. III, we only need the cosine of the angle $\vartheta_f(t)$ between ${\bf e}_f(t)$ and the normal ${\bf z}(t)$ to the pick--up loop plane to study the trapped field signal; together with the loop, ${\bf z}(t)$ rotates about $\hat\omega_r$ with the frequency $\omega_r$ (see fig. S2): 
$$
{\bf z}(t)=\cos({\pi/2}-\alpha){\hat\omega_r}+\sin({\pi/2}-\alpha)\bigl(\cos\theta_r\,{\bf x}_r+\sin\theta_r\,{\bf y}_r\bigr)\equiv\sin\alpha\,{\bf z}_r+\cos\alpha\,\bigl(\cos\theta_r\,{\bf x}_r+\sin\theta_r\,{\bf y}_r\bigr)
$$
\begin{equation}
\theta_r=\theta_{r}(t)=\omega_{r}t=roll\,\,phase
\label{eq:z(t)}
\end{equation}
By means of this,~(\ref{eq:ef}) and formulas~(\ref{eq:lrrelations}) relating the $r$-- and $L$--coordinates, to the first order in the misalignments $\beta_0$ and $\alpha$ we obtain (quadratic and higher order terms are several orders below the required GP-B accuracy):
$$
\cos\vartheta_f(t)=a_{s-r}\sin\Theta_{s-r}+a(\beta_0 \sin\theta_r+\alpha)
$$
\begin{equation}
\Theta_{s-r}(t)=(\omega_s-\omega_r)t+q_{s-r};\qquad \theta_r(t)=\omega_rt
\label{eq:cosTflin}
\end{equation}
For a perfectly spherical rotor $\Delta I=0$ and the amplitudes and initial phase here are true constants whose values depend only on the position of a fluxon relative to the symmetry axis, $a_{s-r}=\sin\xi,\quad q_{s-r}=\eta,
\quad a=\cos\xi$. If, on the other hand, $\Delta I\not=0$, they start to vary slowly with the time at the polhode frequency  according to
$$
a_{s-r}(\omega_pt)=\sqrt{\bigl[\cos\xi\sin\gamma_B+\sin\xi\cos(\omega_pt+\theta_p^0+\eta)\bigr]^2+\sin^{2}\xi\cos^{2}\gamma_B\sin^{2}(\omega_pt+\theta_p^0+\eta)}
$$
$$
\tan q_{s-r}(\omega_pt)=\frac{\sin\xi\cos\gamma_B\sin(\omega_pt+\theta_p^0+\eta)}{\cos\xi\sin\gamma_B+\sin\xi\cos(\omega_pt+\theta_p^0+\eta)}
$$
\begin{equation}
a(\omega_pt)=\cos\xi\cos\gamma_B-\sin\xi\sin\gamma_B\sin(\omega_pt+\theta_p^0+\eta)
\label{eq:amplitudes}
\end{equation}

Note that under the conditions of the GP-B experiment the spin frequency is always much larger than the roll and polhode ones, $\omega_r\sim 5\times10^{-5}\omega_s$, $\omega_p\sim 10^{-5}\omega_s$. Since generally the second term in the first of equations (21) is about five orders of magnitude smaller than the first one, the input signal for the trapped flux output $\Phi_f(t)=\left(\Phi_0/2\right)F_\delta(\cos\vartheta_f(t))$ is a single carrier harmonics of the (high) spin minus roll frequency ($\Theta_{s-r}$), slowly modulated in the phase and amplitude at polhode frequency, added by a small D.C. offset ($\alpha a$), and a small low frequency harmonics ($\theta_r$), both modulated at $\omega_p$. 
Therefore it is natural and convenient to represent $\Phi_f(t)$ as a Fourier series of spin minus roll harmonics with the amplitudes modulated by low frequencies, namely:
$$
\Phi_f(t)=\frac{\Phi_0}{2}\,F_{\delta}(\cos\vartheta_f(t))=
$$
$$
\frac{\Phi_0}{2}\,
\biggl[
a_{s-r}(\omega_p t)\,\sum_{k=0}^\infty\,
A_k(\omega_p t)\,\sin(2k+1)\Theta_{s-r}(t)+a(\omega_p t)\,(\beta_0\sin\omega_r t+\alpha)\,\sum_{k=0}^\infty\,B_k(\omega_p t)\,\cos2k\Theta_{s-r}(t)
\biggr];
$$
$$
A_k(\omega_p t)={2\over\pi(2k+1)}\int_0^\pi\,\cos(2k+1)\psi\,\cos\psi\,
F_\delta^{\prime}(a_{s-r}(\omega_p t)\sin\psi)\,d\psi+O(\beta_0^2);
$$
\begin{equation}
B_k(\omega_p t)={2\over\pi(1+\delta_{k0})}\int_0^\pi\,\cos2k\psi\,
F_\delta^{\prime}(a_{s-r}(\omega_p t)\sin\psi)\,d\psi+O(\beta_0^2);
\label{eq:1fFourierslow}
\end{equation}
here prime denotes the derivative of $F_\delta(s)$ in $s$.

As readily seen, the amplitudes of {\it odd} harmonics of $\Theta_{s-r}$ ($A_k$) are generally of the order of unity and decrease as $O(k^{-2})$ for the large enough number $k$. In contrast with that, the amplitudes of {\it even} harmonics, which are linear in the misalignments, are at least four orders of magnitude smaller but decrease only as $O(k^{-1}),\,k\rightarrow\infty$. In addition, the even harmonics are modulated also by the roll frequency $\omega_r$, so that, along with the harmonics $2k\Theta_{s-r}(t),\quad k=0,1,\dots,$ with amplitudes $\alpha\, a_r(\omega_p t)\,B_k(\omega_p t)$, harmonics  $2k\Theta_{s-r}(t)\pm \omega_r t$ are present, whose amplitudes differ only by the misalignment involved, $0.5\beta_0$ instead of $\alpha$.

With all this in mind, one can easily understand that the full spectrum of the trapped flux signal consists of the following series of frequencies: $(2k+1)(\omega_s-\omega_r)\pm m\omega_p,\quad 2k(\omega_s-\omega_r)\pm \omega_r\pm m\omega_p$ and  $2k(\omega_s-\omega_r)\pm m\omega_p,\quad m,k=0,1,\dots$. The highest peaks are at $(2k+1)(\omega_s-\omega_r)$, and those at $2k(\omega_s-\omega_r) \pm \omega_r$ and $2k(\omega_s-\omega_r)$ are four to five orders of magnitude smaller. All of them are surrounded by an appropriately scaled forest of side bands separated by $\pm m\omega_p$. 

The only remaining thing is to discuss briefly the total flux $\Phi$ produced by {\it all} fluxons. There are always some $N$ pairs of fluxons and {\it antifluxons} present on the rotor's surface after cooling the rotor down below the transition temperature (the antifluxon is a fluxon with the opposite sign of the magnetic field). Experiments have indicated that the expected number of the pairs is around $N\sim 100$, at the most. We denote any  values related to either fluxons or antifluxons by indices $f$ and $a$, respectively, numbering them with the index $i=1,2,\dots,N$; for instance, their body coordinates will be
$\xi_f^i,\,\eta_f^i$ and $\xi_a^i,\,\eta_a^i$, the input signals $S_f^i(t)=\cos\vartheta_f^i(t),\,S_a^i(t)=\cos\vartheta_a^i(t)$, etc.

The general expression for the total trapped field flux is given by
\begin{equation}
\Phi(t)=\sum_{i=0}^N\Bigl[\Phi^i_f(t)+\Phi^i_a(t)\Bigr]
={\Phi_0\over 2}\,\sum_{i=0}^N\,\Bigl[F_{\delta}(\cos\vartheta^i_f(t))-F_{\delta}(\cos\vartheta^i_a(t))\Bigr];
\label{eq:fluxtot}
\end{equation}
obviously, the full spectral representation of $\Phi(t)$ is just a scaled up version of $\Phi_f(t)$ given in~(\ref{eq:1fFourierslow}).

Since for small $\delta$ the transfer function $F_{\delta}(s)$ is close to $\pm F_{\delta}(1)\approx\pm1$ everywhere except a small vicinity of the origin (see sec. III), expressions~(\ref{eq:fluxtot}),~(\ref{eq:1fFourierslow}) demonstrate that the maximum value of $\Phi(t)$ is distributed according to the usual counting statistics, provided that the distribution of fluxons over the surface of the rotor is the uniform random one. Therefore $N$ fluxon-antifluxon pairs in this case should produce the total flux on the order of $\sqrt{N}\Phi_0$ for `large' $N$.

\section{Code and signal analysis}

For the GP-B error analysis and data reduction one needs to simulate the trapped flux signal as expected in the SQUID output. To do that, the results obtained in the previous sections were utilized for writing a program able to fast enough generate, store, and analyze the high-frequency signal. The code  written in the {\it MatLab v.5.0}, to ensure compatibility with other GP-B software, is available from the authors.

The program is very versatile, allowing for many options and many different tasks. For instance, there may be a different {\it number of fluxons}, their {\it positions} may be read either from a prewritten file or generated at random according to different probability distributions. {\it Transfer function} may be calculated by means of several different expressions introduced in sec. III. Generation of the {\it high frequency signal} and/or its slow varying {\it Fourier amplitudes}~(\ref{eq:1fFourierslow}),~(\ref{eq:fluxtot}) is possible. In addition, all gyroscope and pick-up loop parameters (radii, rotor asphericity, misalignments, etc.), as well as the discretization frequency, time intervals, and all angular velocities may be specified in an arbitrary way.

A lot of attention in the program's realization has been paid to the fact that tracing positions of  as much as $200$ fluxons for long enough periods of time with high discretization frequency easily becomes too memory consuming. The program thus has been optimized in several directions, such as not to cause excessive memory swaps to the hard drive, not to lead to the memory fragmentation, and to access the hard drive for data storage as infrequently as possible. The following data may be useful to estimate the code's speed: on a {\it Sun UltraSparc} 5 with 128 Megabytes of RAM running {\it System V, Rel. 4.0} and having a network mounted storage drive it takes, depending on the network load, 1.5 up to 2 hours to generate one hour of signal of $100$ fluxon pairs at a sampling frequency of $2200 Hz$ (the actual sampling rate of GP-B electronics).

Here we will not elaborate more on the code details but continue with the results of our simulations. All of them have been performed with the parameters set at the values expected for the GP-B experiment (see c. f.~\cite{turn,buch,muhl}). In particular, the spin frequency $f_{spin}=100\, Hz$, the roll period $T_{r}=3\, min$, the polhode period $T_{p}\approx 43.6 \, min$; recall that $\delta=0.025$.

In fig. 5 the signals are seen generated by different number of fluxons distributed in various ways over the surface of the gyroscope. In all of the graphs the 'adjusted arctangent' approximation~(\ref{eq:univ_atan1}) to the universal curve is used.  Fig. 5,a shows signals of a single fluxon (without an antifluxon counterpart) positioned at different points on the gyro. The majority of fluxon positions provide signals like the one drawn in the solid line in the figure. The dashed and dash-dotted lines correspond to rare fluxons oscillating in the small ($\sim \Delta_{\delta}$) vicinity of the pick-up loop plane, which is why their amplitude is smaller. On the average, one cannot expect too many fluxons like that, however, each of the four GP-B rotors will carry just one {\it particular} realization of the fluxon position distribution, so these `weak' fluxons are possible.

Fig. 5,b shows various signals from one fluxon-antifluxon pair. Again, the solid line correspond to `the most probable' signal: fluxon and antifluxon are far from each other (though not opposite on the sphere) and have large oscillation amplitudes.

Fig. 5,c shows typical signals of 5, 15, and 100 pairs distributed {\it randomly} with the {\it uniform} probability over the gyro surface. The $\sqrt{N}$ growth of the signal is visible; the complexity of the signal profile also clearly increases with $N$.
\vskip2mm
\centerline{\epsfxsize=0.8\hsize\epsffile{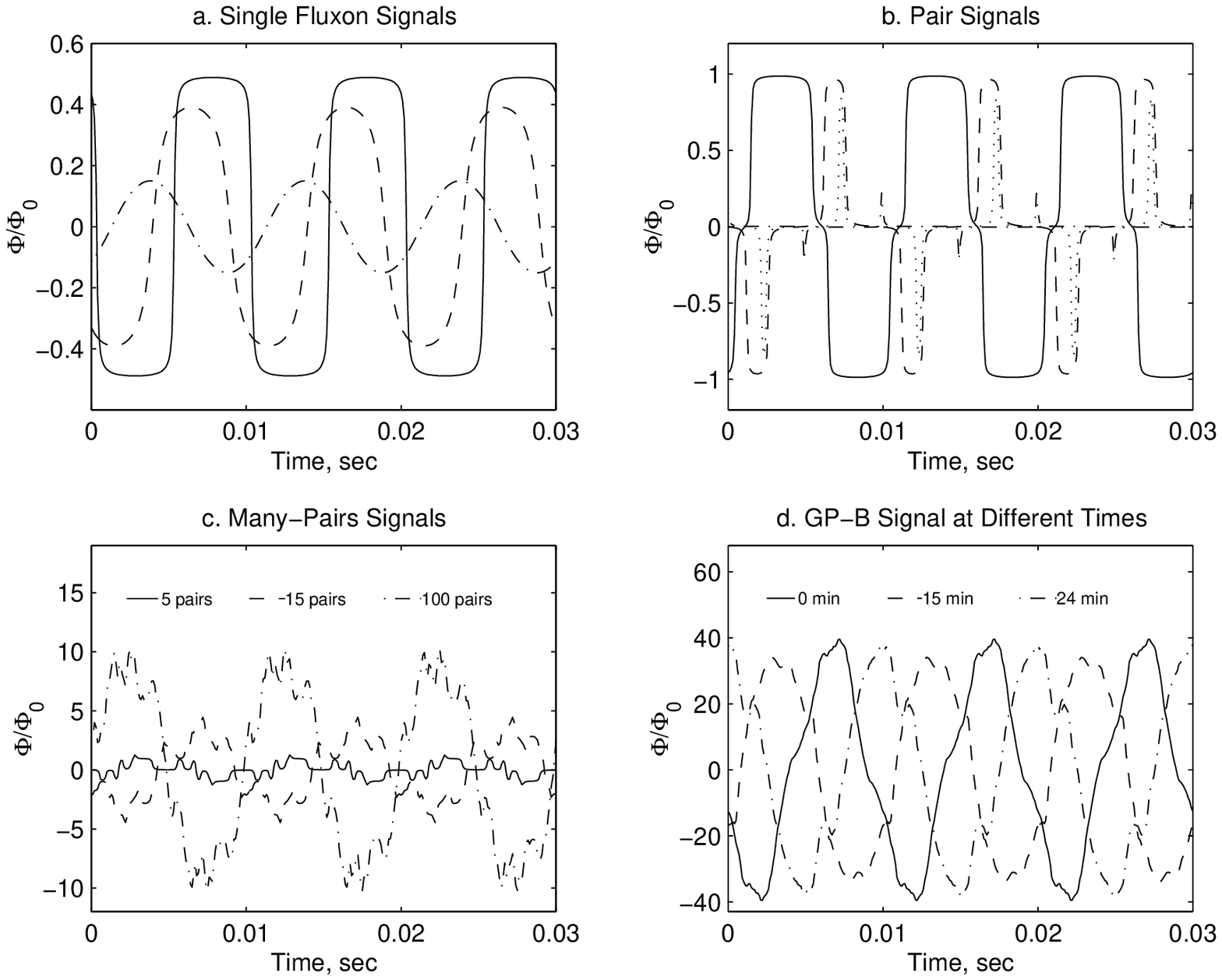}}
\vskip1mm
\centerline{Fig. 5. Simulated Readout Signals.}
\vskip2mm

Fig. 5,d shows short fragments of the 12 hours of signal generated for the test of the GP-B data reduction algorithms. There are 100 fluxon pairs distributed unevenly: 60 of them are uniformly spread at random over the surface (just like in fig. 5,c), while the remaining 40 are used to create a total net flux of $\sim 40 \,\Phi_0$ along some random axis. This should account for a small residual magnetization of the rotor at the time when it was made superconducting (see\cite{PhD}). This magnetization not only significantly increases the amplitude of the signal, but also smoothes it out. Different curves in the figure correspond to the signals taken at different stages of the polhoidal motion (namely, 0, 15, and 24 minutes from some reference point) for a duration of 3 spin periods. 

In fig. 6 a low-frequency envelope is plotted of the signal from fig. 5,d used in GP-B simulations. The graph was constructed by splitting the magnetic flux signal into two-second blocks (4400 data points in each) and plotting the maximum value of the flux for each block. Periodicity of the large scale structures of the envelope with approximately the polhode period of about $43 \, min$ is clear. On the other hand, a comparison of the signal in any two corresponding regions demonstrates that the short scale features, presumably introduced by the roll frequency and other less intensive harmonics, are not repeated precisely every polhode period $T_p$, which is expected because $T_p$ and the roll period $T_r$ are incommensurable.
\vskip1mm
\centerline{\epsfxsize=0.6\hsize\epsffile{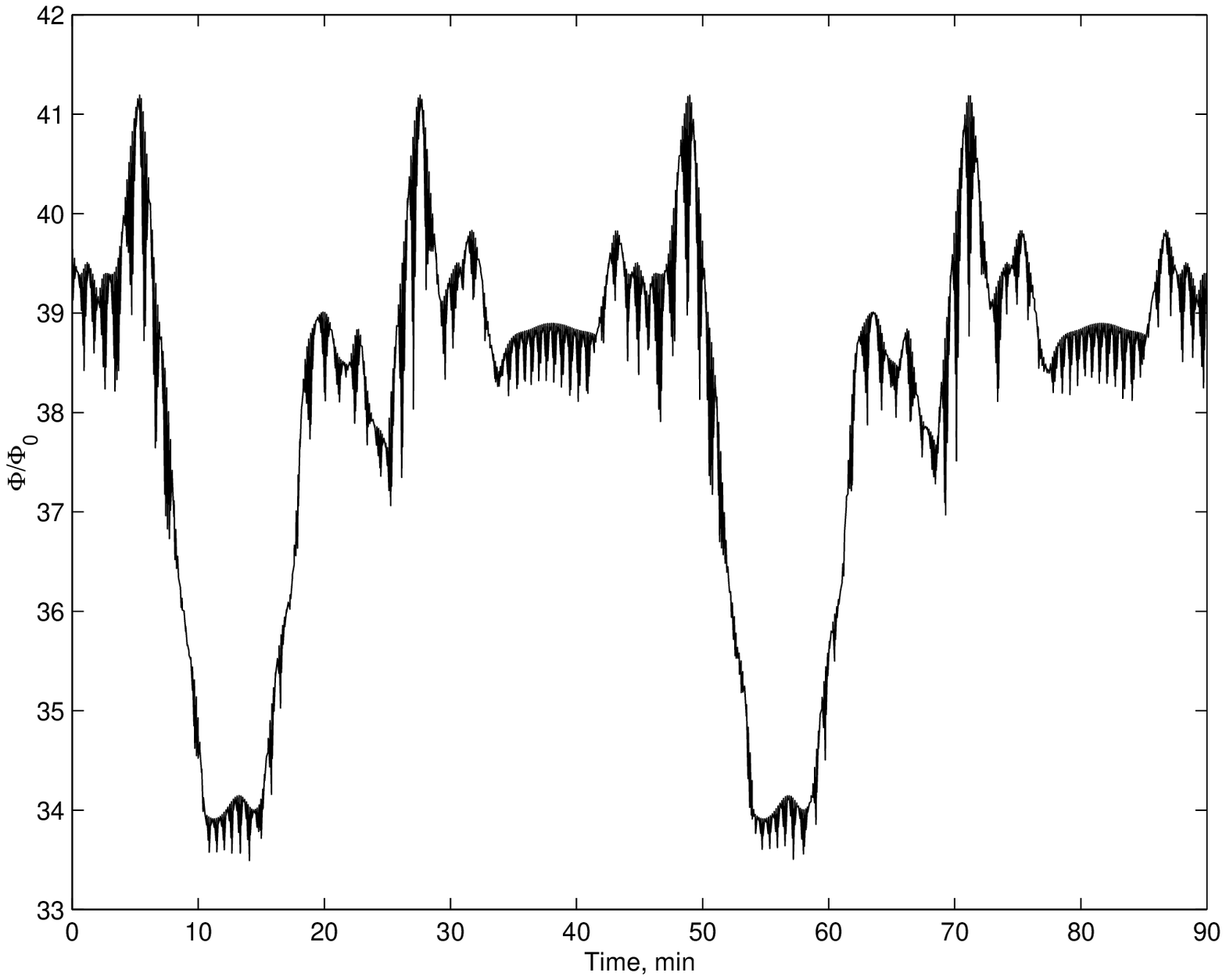}}
\vskip1mm
\centerline{Fig. 6. Envelope of The Simulated Trapped Flux Signal, $T_{p} \approx 43.6 \, min$.}
\vskip2mm
\centerline{\epsfxsize=0.9\hsize\epsffile{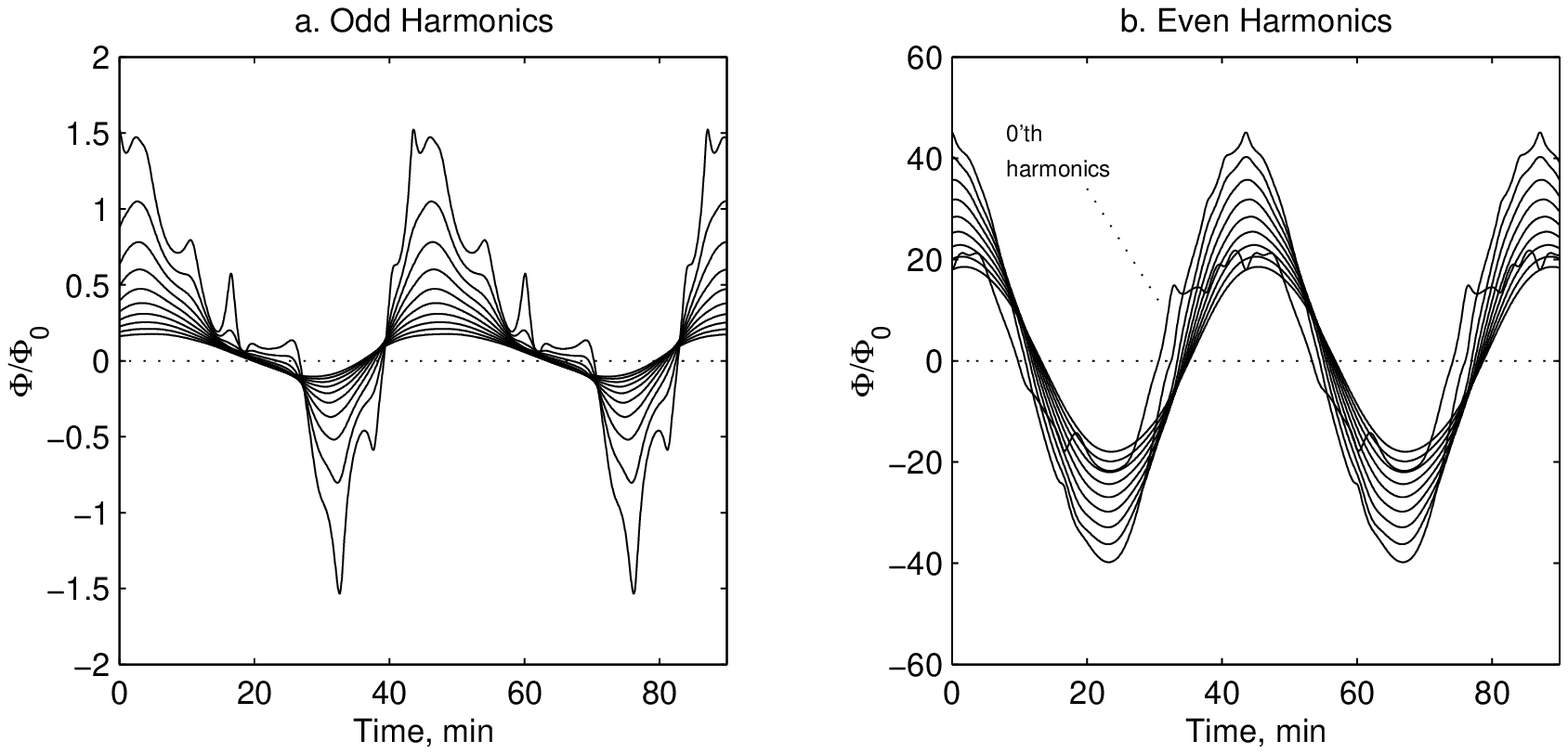}}
\vskip1mm
\centerline{Fig. 7. Slowly Varying Amplitudes of Fourier Harmonics of Trapped Flux Signal, $T_{p} \approx 43.6 \, min$.}
\vskip2mm
Fig. 7 shows the slow polhoidal variation of Fourier amplitudes of the spin minus roll harmonics calculated according to~(\ref{eq:1fFourierslow}) and summed over the fluxons and antifluxons. The first ten odd and even harmonics are shown in plots {\it a} and {\it b}, respectively. Recall that in the expression~(\ref{eq:1fFourierslow}) for the flux all even harmonics are multiplied by the misalignments, so that the actual vertical scale in fig. 7,b is about $ 10^{5}$ of that in fig. 7,a. The pictures clearly show that the odd harmonics drop much faster with the number than the even ones, as predicted. It is interesting to note that the lowest even ($n=0$) harmonics, which gives the amplitude of the D.C. and the roll frequency components, has a shape rather distinctive from the profile of the other modes.

\section*{Acknowledgments}

This work was supported by NASA grant NAS 8-39225 to Gravity Probe~B. We are grateful to Dr. G. M. Keiser, who had originally initiated this work, Dr. M. Heifetz for many valuable comments, and to the Gravity Probe B Theory Group for fruitful discussions.

\section*{Appendix. Summation of Certain Series of Legendre Polynomials}

Here we give a derivation of formulas~(\ref{eq:fd}),~(\ref{eq:kd}) for $f_\delta=F_\delta(1)$ and for the slope $\kappa_\delta$ of the transfer function at $s=0$. We use the Pochgammer symbol $(\alpha)_0=1,\quad (\alpha)_k=\alpha(\alpha+1)\dots(\alpha+k-1)={\Gamma(\alpha+k)/\Gamma(\alpha)}$, as well as the standard notation 
$$
F(a,\,b,\,c;\,\zeta)=\sum_{k=o}^\infty{(a)_k(b_k)\over(c)_k}\,{\zeta^k\over k!}
$$
for the Gauss hypergeometric function of the argument $\zeta$ and parameters $a,\,b,\,c$. From~(\ref{eq:univ_series}) we have
$$
F_\delta(s)\equiv F^{(1)}_\delta(s)-F^{(2)}_\delta(s);
$$
\begin{equation}
F^{(1)}_\delta(s)=2\eta\,\sum_{k=0}^\infty {(-\eta^2)^{k}\over k!}\,\left({1\over2}\right)_k\,P_{2k+1}(s),\qquad
F^{(2)}_\delta(s)={\eta \over 2}\sum_{k=0}^\infty {(-\eta^2)^{k}\over (k+1)!}\,\left({1\over2}\right)_k\,P_{2k+1}(s),
\label{eq:f12}
\end{equation}
where we introduced $\eta=1-\delta$ for brevity.
\vskip2mm
\noindent {\sl Calculation of} $f_\delta$\hskip 1mm. Since $P_n(1)=1$, we have
$$
F^{(1)}_\delta(1)=2\eta\,\sum_{k=0}^\infty {(-\eta^2)^{k}\over k!}\,\left({1\over2}\right)_k= {2\eta\over\sqrt{1+\eta^2}};
$$
$$
F^{(1)}_\delta(1)={\eta \over 2}\sum_{k=0}^\infty {(-\eta^2)^{k}\over (k+1)!}\,{({1\over2})_k\,(1)_k\over(2)_k}={\eta \over 2}\,F(1/2,\,1,\,2;\,-\eta^2)=
\eta^{-1}\left(
\sqrt{1+\eta^2}-1
\right),
$$
and for the elementary expression of the hypergeometric function we have used formula (11) from~\cite{bat1}, 2.11. with $a=1/2,\quad b=1$. Combining these results with~(\ref{eq:f12}), we obtain
$$
f_\delta=F_\delta(s)=F^{(1)}_\delta(s)-F^{(2)}_\delta(s)=
{2\eta\over\sqrt{1+\eta^2}}-{\sqrt{1+\eta^2}-1\over\eta}=
{1\over\eta}\,\left(
1-{1-\eta^2\over\sqrt{1+\eta^2}}
\right),
$$
which, in view of $\eta=1-\delta$, is exactly the expression~(\ref{eq:fd}).
\vskip2mm

\noindent {\sl Calculation of} $\kappa_\delta$\hskip 1mm. As (see~\cite{bat1}, 10.10, (12))
$$
P_{2k+1}^{\prime}(0)=(2k+1)\,P_{2k}(0)={(-1)^{k}\over k!}\,
\left(
{3\over 2}
\right)_k,
$$
from~(\ref{eq:f12}) we find:
$$
{\partial F^{(1)}_\delta \over\partial s}\biggl|_{s=0}=2\eta\,\sum_{k=0}^\infty {(-\eta^2)^{k}\over k!}\,\left({1\over2}\right)_k\,P_{2k+1}^{\prime}(0)=2\eta\,\sum_{k=0}^\infty {(\eta^2)^{k}\over k!}\,{({1\over2})_k({3\over2})_k\over(1)_k}=
$$
\begin{equation}
2\eta\,F(1/2,\,3/2,\,1;\,\eta^2)=
{2\eta\over 1-\eta^2}\,F(1/2,\,-1/2,\,1;\,\eta^2)={4\eta\over \pi(1-\eta^2)}{\bf E}(\eta),
\label{eq:df1}
\end{equation}
where ${\bf E}(\eta)$ is the complete elliptic integral of the second kind, and we have exploited the classical relation (see~\cite{bat1}, 2.1.4, (23))
$$
F(a,\,b,\,c;\,\zeta)=\bigl(1-\zeta\bigr)^{c-a-b}\,F(c-a,\,c-b,\,c;\,\zeta),
$$
and the expression for the elliptical integral in terms of the hypergeometric function (see~\cite{bat3}, 13.8):
\begin{equation}
F(1/2,\,-1/2,\,1;\,\eta^2)={2\over \pi}{\bf E}(\eta)
\label{eq:elli}
\end{equation}

Similarly,
$$
{\partial F^{(2)}_\delta \over\partial s}\biggl|_{s=0}={\eta\over 2}\,\sum_{k=0}^\infty {(-\eta^2)^{k}\over (k+1)!}\,\left({1\over2}\right)_k\,P_{2k+1}^{\prime}(0)=
{\eta\over 2}\,\sum_{k=0}^\infty{(\eta^2)^{k}\over k!}\,{({1\over2})_k({3\over2})_k\over(2)_k}=
$$
\begin{equation}
{\eta\over 2}\,F(1/2,\,3/2,\,2;\,\eta^2)={\eta\over 2} \,(-4)\,{d\over d(\eta^2)}F(-1/2,\,1/2,\,1;\,\eta^2)={-4\eta\over\pi}\,{d\over d(\eta^2)}{\bf E}(\eta)=
-{2\over\pi\eta}\left[{\bf E}(\eta)-{\bf K}(\eta)\right],
\label{eq:df2}
\end{equation}
\noindent and here we used the formula for the derivative of the hypergeometric function (see~\cite{bat1}, 2.8, (20)), formula~(\ref{eq:elli}) again, and a formula for the derivative of ${\bf E}(\eta)$ (see~\cite{bat3}, 13.7, (12)); ${\bf K}(\eta)$ is the complete elliptic integral of the first kind.

Equations~(\ref{eq:df1}),~(\ref{eq:df2}) now provide
$$
\kappa_\delta={\partial F_\delta\over\partial s}\biggl|_{s=0}=
{\partial F^{(1)}_\delta \over\partial s}\biggl|_{s=0}-
{\partial F^{(2)}_\delta \over\partial s}\biggl|_{s=0}=
{2\over\pi\eta}\left[{1+\eta^2\over1-\eta^2}{\bf E}(\eta)-{\bf K}(\eta)\right],
$$
which coincides with the exact expression in~(\ref{eq:fd}); the asymptotic formula there for small $\delta=1-\eta$ is obtained by using the expansions of elliptic integrals in the series in the conjugate modulus (see~\cite{dw}, 773.3, 774.3).

\end{document}